\documentclass[twocolumn]{aastex63}

\received{December 25, 2020}
\revised{March 25, 2021}
\accepted{March 26, 2021}
\submitjournal{ApJ}

\shorttitle{On the origin of the asymmetry of G350.1$-$0.3}
\shortauthors{Tsuchioka et al.}
\graphicspath{{./}{figures/}}
\usepackage{longtable}
\usepackage{multirow}
\usepackage{threeparttable}
\usepackage{color}
\usepackage{lineno}
\begin{document}

\title{On the origin of the asymmetry of the ejecta structure and explosion of G350.1$-$0.3}

\correspondingauthor{Tomoya Tsuchioka}
\email{t.tsuchioka@rikkyo.ac.jp}

\author[0000-0002-8604-1641]{Tomoya Tsuchioka}
\affil{Department of Physics, Rikkyo University, 3-34-1 Nishi Ikebukuro, Toshima-ku, Tokyo 171-8501, Japan}
\author{Yasunobu Uchiyama}
\affil{Department of Physics, Rikkyo University, 3-34-1 Nishi Ikebukuro, Toshima-ku, Tokyo 171-8501, Japan}
\affil{Graduate School of Artificial Intelligence and Science, Rikkyo University, 3-34-1 Nishi Ikebukuro, Toshima-ku, Tokyo 171-8501, Japan}
\author[0000-0001-6409-7735]{Ryota Higurashi}
\affil{Department of Physics, Rikkyo University, 3-34-1 Nishi Ikebukuro, Toshima-ku, Tokyo 171-8501, Japan}
\author[0000-0003-4505-8479]{Hiroyoshi Iwasaki}
\affil{Graduate School of Artificial Intelligence and Science, Rikkyo University, 3-34-1 Nishi Ikebukuro, Toshima-ku, Tokyo 171-8501, Japan}
\author[0000-0002-1644-3266]{Shumpei Otsuka}
\affil{Department of Physics, Rikkyo University, 3-34-1 Nishi Ikebukuro, Toshima-ku, Tokyo 171-8501, Japan}
\author[0000-0003-4808-893X]{Shinya Yamada}
\affil{Department of Physics, Rikkyo University, 3-34-1 Nishi Ikebukuro, Toshima-ku, Tokyo 171-8501, Japan}

\author[0000-0001-9267-1693]{Toshiki Sato}
\affil{Department of Physics, Rikkyo University, 3-34-1 Nishi Ikebukuro, Toshima-ku, Tokyo 171-8501, Japan}
\affil{RIKEN, 2-1 Hirosawa, Wako, Saitama 351-0198, Japan}
\affil{NASA, Goddard Space Flight Center, 8800 Greenbelt Road, Greenbelt, MD 20771, USA}
\affil{Department of Physics, University of Maryland Baltimore County, 1000 Hilltop Circle, Baltimore, MD 21250, USA}



\begin{abstract}

We present X-ray analysis of the ejecta of supernova remnant G350.1$-$0.3 observed with \textit{Chandra} and \textit{Suzaku}, 
and clarify the ejecta's kinematics over a decade and obtain a new observational clue to understanding the origin of the asymmetric explosion.
Two images of \textit{Chandra} X-ray Observatory taken in 2009 and 2018 are analyzed in several methods, 
and enable us to measure the velocities in the plane of the sky. 
A maximum velocity is 4640$\pm$290 km s$^{-1}$ (0.218$\pm$0.014 arcsec yr$^{-1}$) in the eastern region in the remnant. 
These findings trigger us to scrutinize the Doppler effects in the spectra of the thermal emission, 
and the velocities in the line-of-sight direction are estimated to be a thousand km s$^{-1}$. 
The results are confirmed by analyzing the spectra of \textit{Suzaku}. 
Combining the proper motions and line-of-sight velocities, the ejecta's three-dimensional velocities are $\sim$3000--5000 km s$^{-1}$.
The center of the explosion is more stringently constrained by finding the optimal time to reproduce the observed spatial expansion. Our findings that the age of the SNR is estimated at most to be 655 years, and the CCO is observed as a point source object against the SNR strengthen the ''hydrodynamical kick'' hypothesis on the origin of the remnant.

%the ejecta's proper motion is $\sim$4000 km s$^{-1}$ (the distance to the object was assumed to be 4.5 kpc from \cite{2008ApJ...680L..37G}). 
%In terms of line-of-sight, redshift velocities of $\sim$500--1500 km s$^{-1}$ from \textit{Chandra}'s data and $\sim$1500 km s$^{-1}$ from \textit{Suzaku} was measured in the same eastern bright region. These results suggest that the source is young (about 655 years old) and is an asymmetric supernova remnant. In this analysis, we were able to get closer to not only the distribution but also the kinetics of the explosion, which may be an important sample to understand the cause of the asymmetric explosion.

\end{abstract}



\section{Introduction} \label{sec:intro}

The explosion mechanism leading to core-collapse supernovae (CC SNe) is a long-standing problem in modern astrophysics. 
Asymmetry has played an important role in driving the CC SNe, where 
the pattern of the ejecta distribution of supernova remnants (SNRs) is a unique tool to trace the history back to the explosion \citep[e.g.,][]{2000ApJ...528L.109H,2009ApJ...706L.106L}. 
In particular, recent observational studies indicate that understanding the relation between asymmetric ejecta distribution and a motion of the neutron star (NS), so-called ``NS kick'', is critical to reveal the mechanism of the supernova \citep[e.g.,][]{2017ApJ...844...84H,2018ApJ...856...18K}. 
However, there are very few SNRs that allow us to investigate the kinematics of both ejecta and NS directly, 
making it challenging to compare the observation with the theory on the supernova models.

G350.1$-$0.3 is known to be a galactic supernova remnant with a strikingly non-spherical shape of the X-ray image. 
XMMU J172054.5$-$372652 is suggested as a candidate of the ``central compact object (CCO)'' of the remnant, 
thus indicating that the ejecta structure is biased to the east from the putative center of the explosion \citep{2008ApJ...680L..37G,2011ApJ...731...70L}. 
They argue that the remnant is very young (600--1200 yr) and has a transverse motion of the CCO with a relatively high speed of 1400--2600 km s$^{-1}$ for an assumed center of the explosion. 
The estimated age and the CCO speed offer the possibility that the ejecta and the CCO kinematics can be observationally constrained with the X-ray spectroscopy and imaging of the existing X-ray missions. \cite{2020arXiv201112977B} discovered that there is a variation in the brightness distribution in X-rays in several regions of G350.1$-$0.3. They used the maximum likelihood method to measure sky-plane velocities, which were 5000 km s$^{-1}$ for the ejecta and 320 km s$^{-1}$ for the NS.
They also estimated the line-of-sight velocity from spectral analysis to be 900--2600 km s$^{-1}$ redshift. They conclude that the age of G350.1$-$0.3 is younger than 600 years from the fast expansion proper motion.

In this paper, we present the measurements of the ejecta velocities in the three-dimensional (3D) space and the proper motion of the CCO in G350.1$-$0.3 using \textit{Chandra} \citep{2000SPIE.4012....2W} observations in 2009 and 2018. The three independent measurements of proper motion, (1)the optical flow, revealing the overall movement in the SNR, (2)the projection analysis, making it visually more clear, (3)the maximum likelihood method to quantify the movement, performed in this study consistently suggest that the speeds of the ejecta are ~2000-5000 km s$^{-1}$. Furthermore, the finite radial motion is confirmed by the reanalysis of the \textit{Suzaku} \citep{2007PASJ...59S...1M} data.
Our results support a ``hydrodynamic kick'' mechanism based on the firm evidence that NS is kicked to the opposite direction of the spread of the ejecta, 
suggesting that the remnant can be the unique sample to investigate the origin of the asymmetry in CC SNe. Throughout this paper, uncertainty intervals of 90\% confidence level are quoted unless explicitly stated otherwise.

\section{Observations} \label{sec:obs}

The \textit{Chandra} X-ray Observatory has observed G350.1$-$0.3 in 2009 April (PI: P. Slane) and 2018 July (PI: S. Reynolds) using the Advanced CCD Imaging Spectrometer Spectroscopic-array (ACIS-S). 
The total exposure times of the two observations are 82.97 ks and 189.24 ks, respectively. More detailed information on individual IDs is shown in table 1.
The data are reprocessed using {\tt chandra\_repro} in CIAO 4.11 with CALDB 4.8.2, provided by the \textit{Chandra} X-Ray Center.

In order to improve the positional accuracy of the images, eight point sources are identified using {\tt wavdetect} and the coordinate difference between the two epochs is evaluated by {\tt wcs\_match}. 

Then the coordinates of the event files are updated with {\tt wcs\_update} to minimize the differences. 
The correction results in the average positional accuracy of 0.378$^{\prime\prime}$.  
Using eight point sources, this value was calculated as follows
\begin{equation}
\overline{\delta d}=\sqrt{\frac{\sum_{i=1}^{N} \delta d_{i}^{2}}{N}}
\end{equation}
where $\delta d$ is the distance between the coordinates in 2009 (updated by using {\tt wcs\_update}) and 2018, $N$ is the number of point sources. It gives a reference to the systematic uncertainties on the positional accuracy. 

The X-ray image of G350.1$-$0.3 taken in 2009 is shown in figure 1. There is little or no X-ray emission to the west, and XMMU J172054.5-372652 is observed \citep{2011ApJ...731...70L}. It contrasts to typical supernova remnants, such as a spherical morphology seen in Tycho's SNR. 
The X-ray intensity is higher in the eastern region than in the northern and southern one. 
XMMU J172054.5$-$372652 is located in the western region, which about 2$'$ away from the bright eastern region. 

XMMU J172054.5--372652 is observed in 2013 May with ACIS Continuous Clocking (CC) Mode (PI: Gotthelf), listed in table 1 for more details.
The CC Mode is different from the Very Faint (VF) Mode used in the other G350.1$-$0.3 \textit{Chandra} observations in that it sacrifices spatial resolution in two dimensions, so this data cannot be used for image analysis.
The data are used only for CCO spectrum analysis.

The \textit{Suzaku} X-ray Observatory has observed G350.1$-$0.3 in 2011 September \citep{2014PASJ...66...68Y}. 
The data from the front-illuminated CCDs \citep[XIS0 and XIS3;][]{2007PASJ...59S..23K} are used 
and reprocessed with the current calibration database (2016 February 14) using the standard screening criteria. 
The effective exposure time is 70.2 ks. 
The X-ray spectrum is extracted from a circle region with a 1.5$^{\prime}$ radius that covers the bright eastern structure. 
The background spectrum is extracted from the dim spot in the same data, which is a typical way of estimating the background.

\begin{deluxetable*}{ccccccccc}[ht!]
\tablecaption{Observational log of the \textit{Chandra} and \textit{Suzaku} data.}
\tablehead{
 \colhead{Observatory} & \colhead{Target Name} & \colhead{Obs. ID} & \colhead{Obs. Start} & \colhead{Exposure} & \colhead{Detector\tablenotemark{a}}& \colhead{RA}& \colhead{Dec}& \colhead{Roll}\\ 
\colhead{} & \colhead{} & \colhead{} &\colhead{(yyyy mm dd)} &\colhead{(ks)} &\colhead{} & \colhead{(deg)} & \colhead{(deg)} & \colhead{(deg)}
}
\startdata
\hline
\textit{Chandra} & G350.1$-$0.3& 10102&  2009 Apr 21& 82.97&ACIS-S (VF)& 260.2653& -37.4388& 57.9084\\
& G350.1$-$0.3& 20312& 2018 Jul 2& 40.58& ACIS-S (VF)& 260.2615& -37.4476& 296.1505\\
& G350.1$-$0.3& 20313&  2018 Jul 4& 19.84& ACIS-S (VF)& 260.2606& -37.4470& 296.1504\\
& G350.1$-$0.3& 21118&  2018 Jul 8& 42.91& ACIS-S (VF)& 260.2608& -37.4471& 296.1497\\
& G350.1$-$0.3& 21119 &  2018 Jul 7& 48.29 &ACIS-S (VF)& 260.2605& -37.4468& 296.1490\\
& G350.1$-$0.3& 21120&  2018 Jul 5 & 37.62 &ACIS-S (VF)& 260.2608& -37.4472& 296.1490\\
& XMMU & 14806 & 2013 May 11 & 89.33 &ACIS-S (CC)& 260.2255& -37.4454& 64.2424\\[-1.5mm]%
&J172054.5-372652 \\
\hline
\textit{Suzaku} &G350.1$-$0.3 & 506065010 &  2016 Feb 14& 70.18 & XIS-0, XIS-3& 260.2697& -37.4549& 266.4234\\
\enddata
\tablenotetext{a}{The terms in parentheses indicate the mode of imaging. VF and CC represent the Very Faint Mode and Continuous Clocking Mode of ACIS, respectively.}
\label{tab:chandra_dataset}
\end{deluxetable*}

\section{Results} \label{sec:results}

\subsection{Sky-plane velocities from image analysis}

\begin{figure*}[htb!]
 \centering
 \includegraphics[width=18cm]{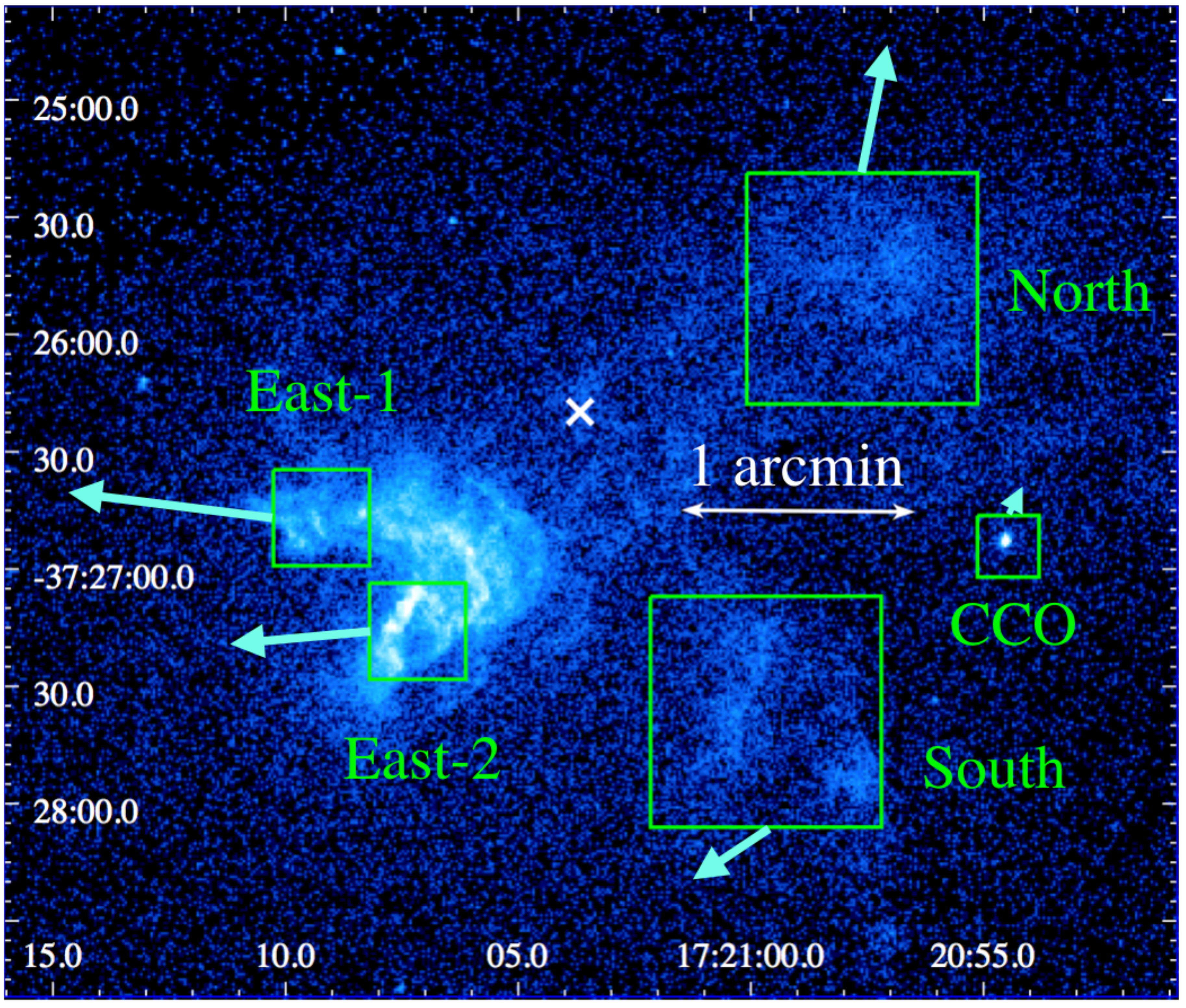}
 \caption{\textit{Chandra} ACIS-S flux image taken in 2009. The energy range is 0.5--7.0 keV.
The five green boxes refer to the proper motion analysis regions using the maximum likelihood method. The cyan arrows indicate the relative magnitude of 2D velocities and directions of motion of the ejecta that we obtained. The white cross marker indicates the on-axis aimpoint for the observation in 2009.}
\end{figure*}

\begin{figure*}[htb!]
 \centering
 \includegraphics[width=18cm]{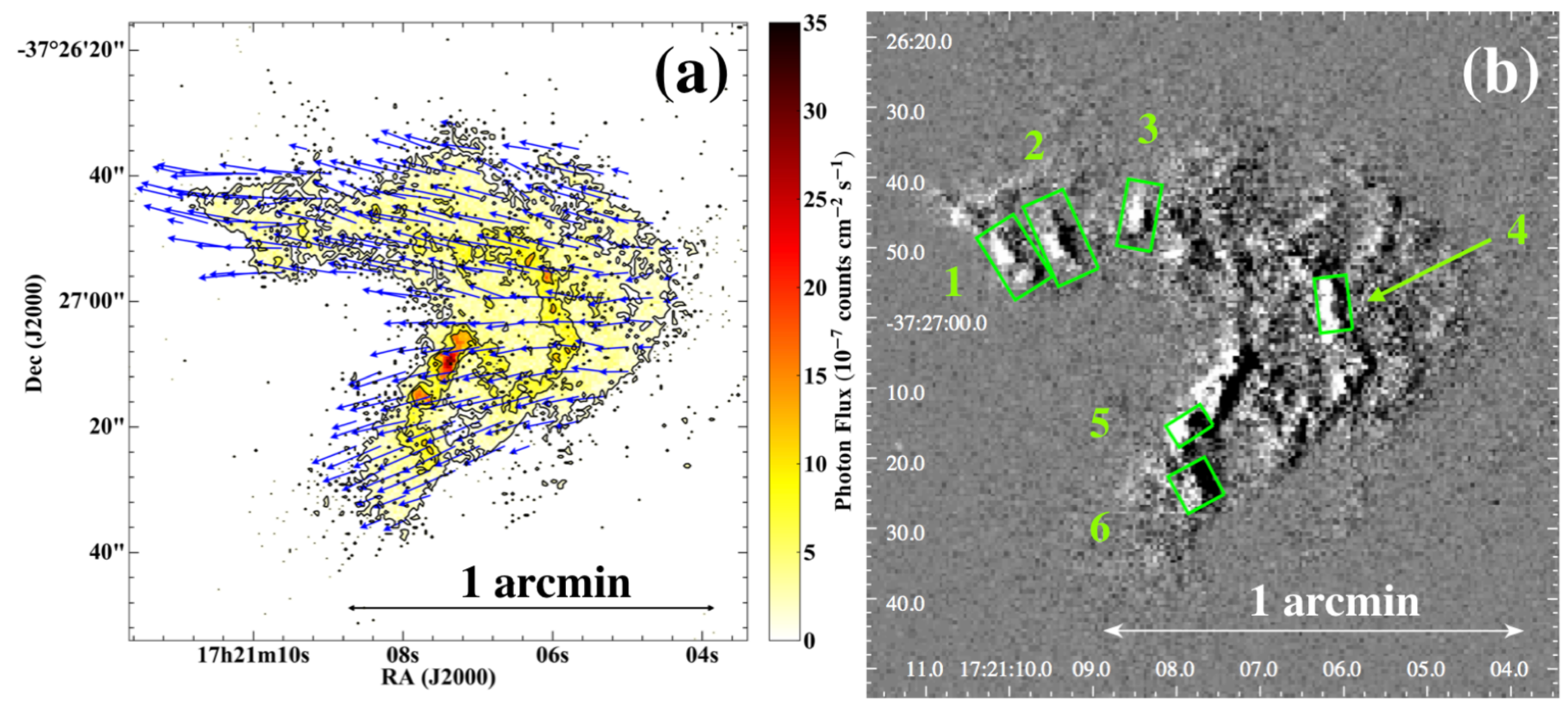}
 \caption{(a) The \textit{Chandra} X-ray image in 2009 overlaid with the vector fields obtained by the optical flow method.
(b) Differential \textit{Chandra} image of G350.1$-$0.3 between 2009 and 2018 observation ($2018-2009$). 
The six green boxes are the regions chosen for proper motion analysis using the projection method. Throughout this figure, the energy range used is 0.5--7.0 keV.}
\end{figure*}

%optical flow
To visually grasp the slight change in the image from 2009 to 2018, 
we adopted one of the dynamic image analyses called ``optical flow''. 
It calculates the flow's two-dimensional vector fields, which obey the conservation law of the total intensity in the image. A similar approach has been applied to Cassiopeia A \citep{sato2018x}. 
More specifically, a derived version of the optical flow called Gunnar Farneb{\"a}ck method \citep{farneback2003two} is 
used to focus more on the small movement of patchy areas. 
The parameters and the setups are the same as used in \cite{sato2018x}. 
When a flux in a pixel is less than $1 \times 10^{-7}$ counts cm$^{-2}$ s$^{-1}$, 
it is regarded as background, and the corresponding pixel is discarded in this calculation. 
The result of the optical flow is shown in figure 2a. 
The obtained vector fields are indicated by the length and the direction of the arrows. 
In this way, we succeeded in visualizing the change in the two images in the bright eastern region, 
and revealing at least two distinctive flows on a large scale: from west to east in the upper half and from west to south-east in the bottom half. 
Note that only vectors with an amplitude greater than $\sim$1 arcsec ($1152 ~\mathrm{km~s}^{-1}$ at the distance of 4.5 kpc) 
are shown in figure 2a, where the counts are significantly higher than the typical level of the background. 

%1D projection
We then proceed in visualizing a change on a much smaller spatial scale. 
The X-ray image of 2009 is subtracted from that of 2018 and shown in figure 2b, 
which works to emphasize the difference between the two images. 
The positive part (white in the figure) means the X-ray intensity in 2018 is brighter than in 2009, and vise versa. 
In other words, the pairs of the black and white region indicate the movement of the image. Filamentary structures are extending from north to south in several places, 
indicating that the ejecta moved in an east-west direction, which is consistent with the results obtained by optical flow. Furthermore, it vividly illustrates a more pronounced but detailed internal structure of the movement in the remnant. 

To further confirm the flow, 
one-dimensional (1D) radial profiles are created from the box regions shown in figure 2b.  
The direction of the flow is defined as the positive values from east to south. The unit is the number of pixels (1 pixel = 0.492$^{\prime\prime}$). 
The projection of the flux is performed by integrating along the direction perpendicular to the flow direction. 
As a result, the projected profiles of the fluxes are obtained from each image of 2009 and 2018. The projection profile from box4 is shown in figure 3 as an example. 
Each profile is fitted with a Gaussian function using the least-squares method. The centers of the Gaussian functions obtained refer to the distance of the move over the two observations, 
resulting in the velocity of the ejecta. 
By assuming that the distance to G350.1$-$0.3 is 4.5 kpc \citep{2011ApJ...731...70L}, 
the proper motion velocities in all the 6 box regions are calculated and summarized in table 2. The ejecta velocities are found to range from 3500 to 5000 km s$^{-1}$. 

\begin{figure}[htb!]
 \centering
 \includegraphics[width=9cm]{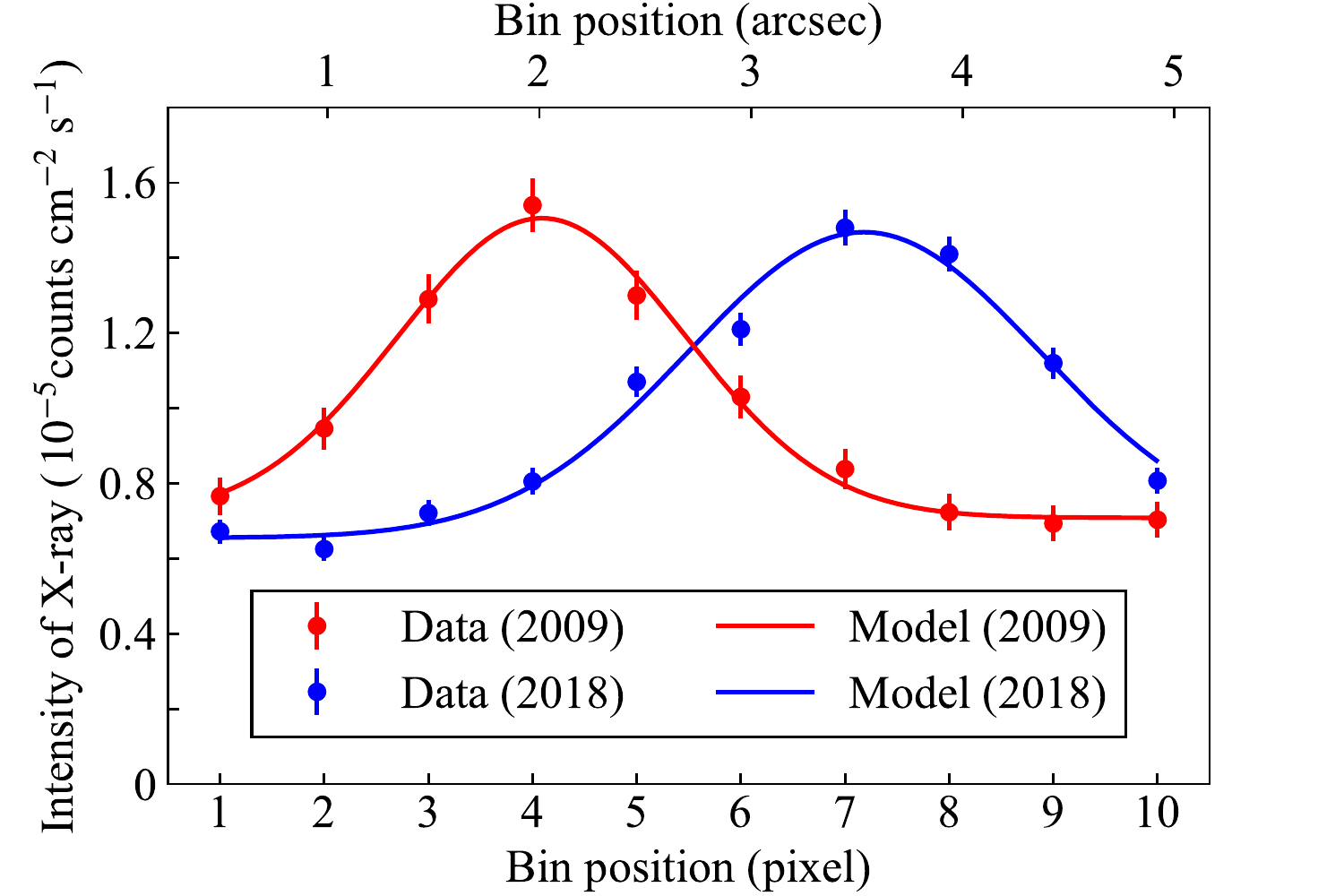}
 \caption{Projection profile created from box4 shown in figure 2b. The red and blue points refer to the data taken in 2009 and 2018, respectively. 
The red and blue curves represent the best-fit Gaussian functions.}
\end{figure}

\begin{deluxetable*}{ccccccc}[ht!]
\tablecaption{Velocity measurements in the regions shown in figure 1, 2b from \textit{Chandra} and \textit{Suzaku} data.}
\tablehead{
 \colhead{region\tablenotemark{a}} & \twocolhead{angular velocity} &  \colhead{line-of-sight velocity} & \colhead{total} & \colhead{\(\chi^{2} \)/d.o.f.} & \colhead{\(\chi^{2} \)/d.o.f.}\\ 
 \cline{2-3}
\colhead{}& \colhead{(arcsec yr$^{-1}$)} & \colhead{(km s$^{-1}$)\tablenotemark{b}}  & \colhead{(km s$^{-1}$)} &\colhead{(km s$^{-1}$) } &\colhead{z=0 (fixed)} &\colhead{z=free}
}
\startdata
\hline
East-1 & $0.218 \pm 0.014$ & $4640 \pm 290$ & $1280 ^{+20}_{-20}$ & $4820 ^{+280}_{-280}$ & $527/251$ & $369/250$\\
East-2   &$0.149 \pm 0.014$ &$3180 \pm 290$ &$1100 ^{+170}_{-190}$ & $3360 ^{+280}_{-280}$ & $760/299$ & $433/298$\\
North & $0.138 \pm 0.027$ & $2940 \pm 580$ & $660 ^{+160}_{-110}$ & $3010 ^{+570}_{-570}$ & $244/252$ & $231/251$ \\
South & $0.097 \pm 0.027$ & $2080 \pm 580$ & $2570 ^{+140}_{-40}$ & $3300 ^{+380}_{-370}$ & $615/219$ & $255/218$ \\
CCO & $0.030 \pm 0.014$ & $640 \pm 290$ & ---  & --- & ---  & --- \\
\hline
box1 & $0.174 \pm 0.028$ & $3720 \pm600$ & $1160 ^{+1400}_{-230}$ & $3900 ^{+710}_{-580}$ & $168/131$ & $134/130$\\
box2 & $0.192 \pm 0.008$ & $4100 \pm170$ & $1310 ^{+700}_{-30}$ & $ 4310 ^{+270}_{-160}$  & $147/132$ & $121/131$\\
box3 & $0.172 \pm 0.015$ & $3670\pm330$ & $1070^{+220}_{-150}$ & $ 3820^{+320}_{-320}$ & $130/114$ & $115/113$\\
box4 & $0.167 \pm 0.009$ & $3650 \pm 200$ & $670^{+250}_{-490}$ & $3620^{+200}_{-220}$  & $167/144$ & $156/143$\\
box5 & $0.181 \pm 0.014$ & $3860 \pm 290$&$470^{+330}_{-100}$&$3890^{+290}_{-290}$   & $136/116$ & $126/115$\\
box6 & $0.227 \pm0.018$ & $4850 \pm390$ & $1540 ^{+300}_{-1210}$ & $5090 ^{+380}_{-520}$  & $148/123$ & $127/122$\\ 
\hline
\textit{Suzaku} East & --- & --- & $1460 ^{+40}_{-10}$ & ---   & $1096/776$ & $599/775$\\
\enddata
\tablenotetext{}{All errors listed in the table represent statistical errors. The systematic errors are discussed in Section 2.}
\tablenotetext{a}{East-1, East-2, North, South, CCO are indicated in figure 1, box1, 2, 3, 4, 5 and 6 are indicated in figure 2b.}
\tablenotetext{b}{The distance to the G350.1$-$0.3 is assumed to be 4.5 kpc. The systematic error obtained from the point source correction described in Section 2 is 884 km s$^{-1}$ by assuming 4.5 kpc as well.}
\label{tab:tab2}
\end{deluxetable*}

% Maximum Likelihood
To estimate the direction of the flow with statistical significance, 
we calculate the 2D proper motions for the box regions in figure 1 by using the maximum likelihood method. 
The method is based on the previous work \citep{2017ApJ...845..167S,sato2018x,2020ApJ...893...98M}. 
The image of 2018 
% the long observation in 2018 no definition 
is used as a template, while that of 2009 is floated to find a 2D vector to maximize the likelihood. 
Here the likelihood function is defined as,
\begin{equation}
L=\prod_{i, j} \frac{\lambda_{i, j}^{k_{i, j}} e^{-\lambda_{i, j}}}{k_{i, j} !},
\end{equation}
where \((i,j)\), \(k_{i, j}\) and \(\lambda_{i, j}\) are the integers to specify the location of the pixel, the photon counts in a pixel \((i,j)\) of 2018, and the photon counts in a pixel \((i,j)\) of 2009, respectively\footnote{To be precise, we multiplied the 2018 count map by the ratio of the exposures map between the two epochs to consider the difference in effective area between them.}. 
By searching the local extrema of the likelihood function by changing the pairs of $i$ and $j$, 
we obtain the significance map in the space of $i$ and $j$. 
It gives the range and the peak of the significance. 
As a result, the ejecta's best-fit transverse velocities are estimated to be $\sim$2000--4600 km s$^{-1}$ as shown in table 2, 
which is consistent with the result of the 1D projection. 
Furthermore, the best-fit 2D vectors are evaluated with their statistical significance. 

% Maximum Likelihood CCO
We apply the same method to the CCO, and obtained its proper motion of $\sim$640 km s$^{-1}$. 
Since the CCO is a point source, 
the systematic uncertainty on the point-spread function (PSF) is not negligible. 
As the roll angle is different between the 2009 and 2018 observations (see table 1), 
the non-axisymmetric pattern of the PSF needs to be taken into account. 
Therefore, we created the simulation images by using a simulator called MARX\footnote{https://space.mit.edu/cxc/marx/}. 
The center of the point source in the observed image is assessed by using the maximum likelihood method with the simulation image. 
By subtracting the corrected coordinate of 2018 from that of 2009, 
the CCO's proper motion between two epochs is evaluated to be $\sim640\pm290$ km s$^{-1}$.

\subsection{Line-of-sight velocities from spectral analysis}
% detailed analysis on the doppler motion
We measure the line-of-sight velocities of the ejecta using the X-ray spectra's Doppler effect. 
We extracted the spectra from four regions (East-1, East-2, North, and South in figure 1) 
and six regions (box1--6 in figure 2b) using the CIAO {\tt specextract} command. 
The energy band of 0.5--6.0 keV was used in the analysis. 

The spectra were fitted with two components in XSPEC (version 12.10.1): a non-equilibrium ionization plasma model (NEI), ``{\tt vvpshock}'', and a photoelectric absorption model with the ``{\tt angr}'' abundance \citep{1989GeCoA..53..197A}. In the fitting procedure, the {\tt\(kT_{e}\)}, {\tt\(n_{e} t\)}, and normalization are treated as free parameters. Abundances of Mg, Al, Si, S, Ar, and Ca are allowed to vary freely, while the other abundances of elements are frozen to the solar values.
The typical value of the equivalent hydrogen column density of the neutral absorber along the line-of-sight, $N_{\mathrm{H}}$, is $3-4 \times 10^{22}\, \mathrm{cm}^{-2}$.
The spectra of the two epochs are combined in {\tt combine\_spectra} because the shape of each data is similar (see Appendix).
Figure 4 shows the spectra of G350.1$-$0.3, the best-fit model, and its residuals for East-1 and East-2 in the energy range of 1.65--2.3 keV. 
Figure 4a shows the fitting result of East-1 when the redshift ($z$) parameter of the NEI model is frozen at zero, 
resulting in \(\chi^{2} \)/d.o.f. of 527/251. The residuals are still present nearby the emission lines, suggesting the shift of the energy scale. 
Therefore, we continue fitting with a floating redshift, and figure 4b shows the fitting results. The residuals significantly decreases (\(\chi^{2} \)/d.o.f. = 369/250). 
By fitting all the spectra with the redshift freed, 
the best-fit values of the redshift are obtained and summarized in table 2.
Si Ly$\alpha$ emission lines ($\sim$2 keV) are visible in the spectra in figure 4. They are the emission lines of which the centroid energy does not depend on the ionization state and are well fitted when $z$ parameters are set to free in this study. By measuring the center and intensity of the Si K$\alpha$ + Ly$\alpha$ lines, we can determine the ion fraction, or ionization state. Now that they are well fitted as well, the ionization state can be determined independently of the doppler effect.
The X-ray plasma in the east, the north, and the south 
are redshifted at 1000--1300 km s$^{-1}$, $\sim$660 km s$^{-1}$, and above 2500 km s$^{-1}$, respectively. 

Since it is commonly challenging to calibrate the energy scale of CCD in orbit, 
it is prudent to confirm the result by using more than one mission. 
We thus analyzed the \textit{Suzaku} data because it has a larger effective area and a higher signal-to-background ratio per arcmin$^2$. 
The spectra are obtained from a circle with a radius of 1.5$^{\prime}$ covering the bright eastern structure. 
As shown in figure 4c and 4d, the residuals become smaller when the redshift parameter is floated. 
The line-of-sight velocity is estimated to be $\sim$1500 km s$^{-1}$. The result is consistent with those obtained with \textit{Chandra}.

\begin{figure*}[htb!]
 \centering
 \includegraphics[width=18cm]{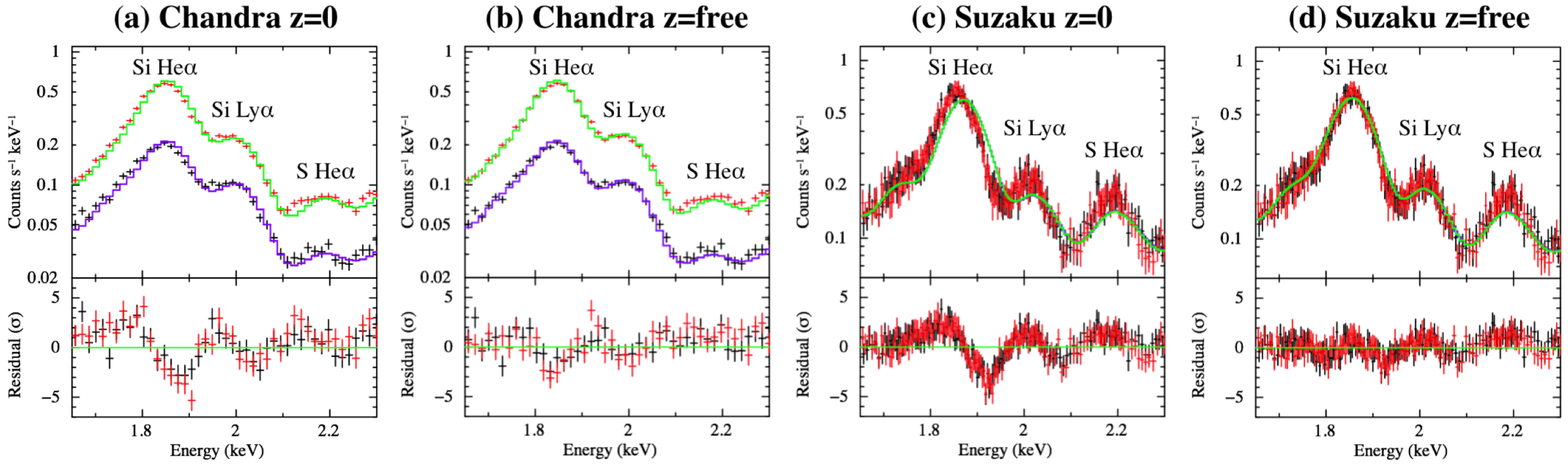}
 \caption{The \textit{Chandra} and \textit{Suzaku} spectra in 1.65--2.3 keV of G350.1$-$0.3. 
(a) The best-fit spectra of \textit{Chandra} and the NEI models (top), and the corresponding residuals (bottom). 
The colors of the data refer to East-1 (black) and East-2 (red). The models are plotted in a solid line. 
(b) The same as (a) except that $z$ is free. (c) and (d) are the same as (a) and (b) except that the data are taken from one region in the \textit{Suzaku} data, 
and the red and black spectra represent the XIS-0 and XIS-3 data.}
\end{figure*}

\subsection{Spectral analysis of XMMU J172054.5--372652}
The CCO spectrum is fitted phenomenologically with the absorbed blackbody radiation model in XSPEC. The energy band of 0.5--6.0 keV is used in the analysis. The spectra of the 2009, 2013, and 2018 observations are individually fitted and then joint-fitted (but only untied for normalization for each epoch).
The results of the best-fit parameters are shown in table 3. The fit is acceptable even without adding extra complex components, and there are no significant changes in parameters between different epochs of observations. The blackbody temperature $kT_{e}$ was also consistent with \cite{2011ApJ...731...70L}. We then fitted the data with a physically-motivated model. According to \cite{2020MNRAS.496.5052P}, the CCO spectrum was fitted with absorbed {\tt nsx} model \citep{2009Natur.462...71H} reproducing the spectrum of the atmosphere from NS. In the analysis of \cite{2020MNRAS.496.5052P}, the 2009 and 2013 observations targeting XMMU J172054.5--372652 (ID: 14806) are also used in the spectral analysis. For better statistics, we add observations from 2018, which are not included in the analysis of \cite{2020MNRAS.496.5052P}. The spectrum of XMMU J172054.5-372652 superimposed on the absorbed {\tt nsx} model is shown in figure 5. Note that there are no significant changes in surface temperature or flux between 2009 and 2018. It can be seen from table 3 that the amount of flux in 2013 with different ACIS modes is larger than that in other epochs. We estimated the amount of pileup in 2009 and 2018 using the tool PIMMS\footnote{https://cxc.harvard.edu/toolkit/pimms.jsp}, and found that it was 13\% and 8\%, respectively. Thus, the pileup may be the reason why the flux in the 2013 observation appears to be the largest in the observations of three epochs.

\begin{deluxetable*}{lccccc}[ht!]
\tablecaption{Best-fit spectral parameters of XMMU J172054.5--372652.}
\tablehead{
 \colhead{Model} &\colhead{Parameter} & \colhead{2009} & \colhead{2018} & \colhead{2013} & \colhead{joint fit}}
\startdata
\hline
absorbed blackbody model\\
{\tt tbabs} &$N_{\mathrm{H}}$ $(10^{22} \, \mathrm{cm}^{-2})$& $3.02 ^{+0.19}_{-0.18}$ & $3.12 ^{+0.15}_{-0.14}$ & $2.70 ^{+0.17}_{-0.16}$& $3.00 ^{+0.10}_{-0.09}$\\
{\tt bbody} &$kT_{e}$ (keV)& $0.52^{+0.02}_{-0.02}$ & $0.51 ^{+0.01}_{-0.01}$ & $0.50 ^{+0.02}_{-0.02}$ & $0.51 ^{+0.01}_{-0.01}$\\
&flux\tablenotemark{a} $(10^{-13} \, \mathrm{ erg \cdot {cm}^{-2} \cdot {s}^{-1}})$& $4.59^{+0.15}_{-0.11}$ & $4.64 ^{+0.08}_{-0.09}$& $5.96 ^{+0.15}_{-0.14}$& ---\\
&\(\chi^{2} \)/d.o.f.& $120/110$ & $173/178$ & $225/202$ & $574/494$\\
\hline
absorbed {\tt nsx} model\\
{\tt tbabs} &$N_{\mathrm{H}}$ $(10^{22} \, \mathrm{cm}^{-2})$& $3.71 ^{+0.23}_{-0.23}$ & $3.89 ^{+0.17}_{-0.18}$ & $3.41 ^{+0.20}_{-0.20}$ & $3.72 ^{+0.12}_{-0.11}$\\
{\tt nsx} &$\log T_{\mathrm{eff}}$ (K)& $6.37^{+0.03}_{-0.03}$ & $6.34^{+0.03}_{-0.02}$ & $6.33^{+0.03}_{-0.03}$ & $6.34^{+0.02}_{-0.02}$\\
&flux\tablenotemark{a} $(10^{-13} \, \mathrm{ erg \cdot {cm}^{-2} \cdot {s}^{-1}})$ & $4.61 ^{+0.09}_{-0.57}$ & $4.64 ^{+0.09}_{-0.28}$ & $5.99 ^{+0.10}_{-0.52}$ & ---\\
&\(\chi^{2} \)/d.o.f.& $112/110$ & $153/178$& $220/202$ & $539/494$
\enddata
\tablenotetext{a}{Flux is calculated in the range of 1.0--6.0 keV.}
\end{deluxetable*}

\begin{figure}[htb!]
 \centering
 \includegraphics[width=8cm]{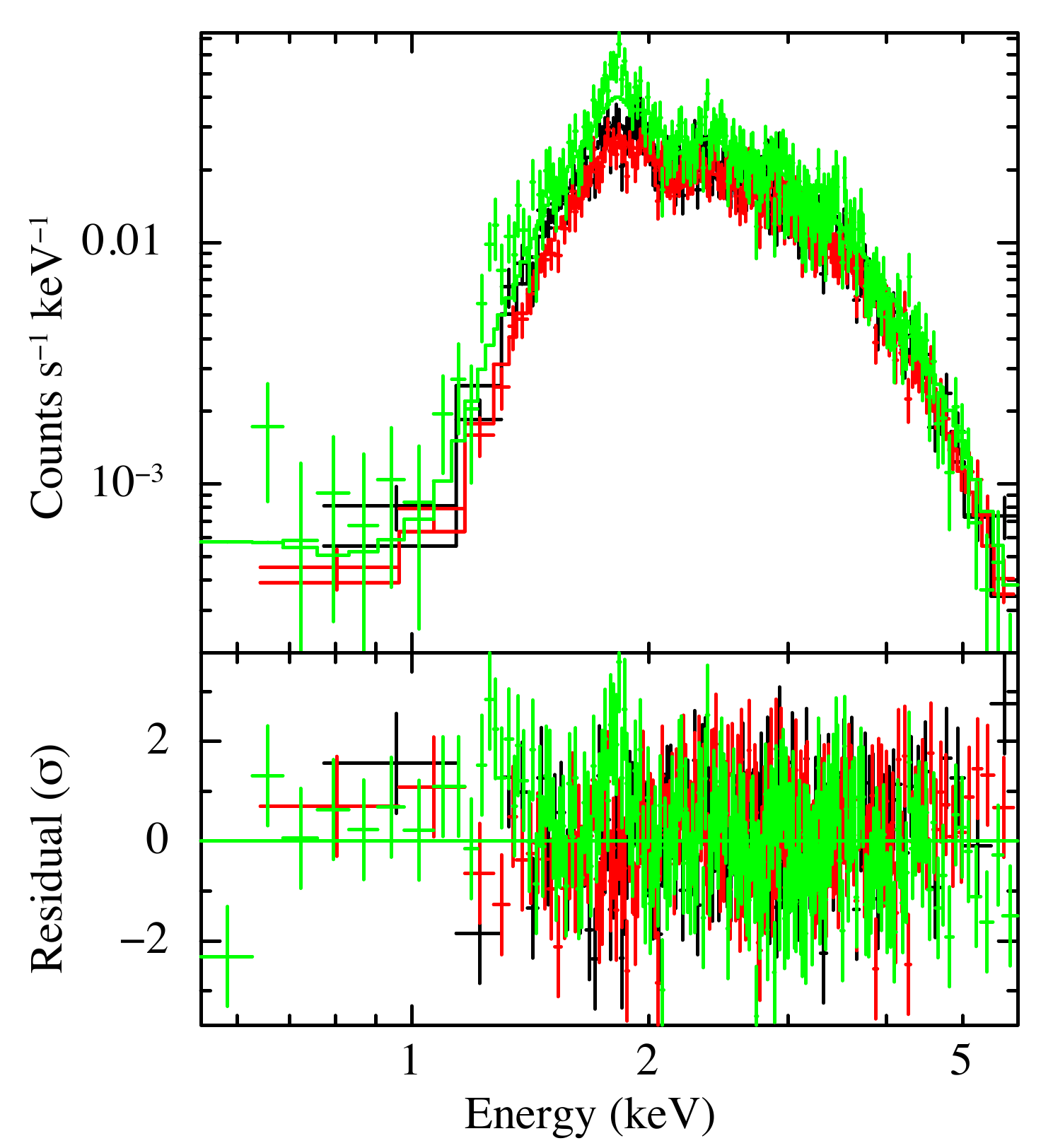}
 \caption{The \textit{Chandra} spectra in 0.5--6.0 keV of XMMU J172054.5$-$372652. The best-fit spectra of \textit{Chandra} and absorbed {\tt nsx} model (top), and the corresponding residuals (bottom). The colors of the data refer to 2009 (red), 2013 (green), and 2018 (black).
 The fit parameters are shown in table 3.}
\end{figure}

\section{Discussion} \label{sec:dis}

We succeeded in measuring the velocities in the plane of the sky of G350.1$-$0.3 through the analysis of the two X-ray images 
owing to \textit{Chandra}'s high spatial resolution. 
By analyzing the flow using several different methods, 
the regions in north, east, and south are found to move in a different direction. 
Thus the velocities in the plane of the sky are regarded as the velocities of the ejecta's proper motion. Inspired by this result, we have attempted to release $z$ to be free, and obtained finite values of $z$ from both \textit{Chandra} and \textit{Suzaku} data. 
This has been overlooked in \cite{2014PASJ...66...68Y}, probably because there was no clear evidence on the proper motion. 
The velocities that we obtained are compiled in table 2. 
These results are generally consistent with the results of \cite{2020arXiv201112977B}, which shows the speed of the G350.1$-$0.3 ejecta.
We reinforced it with different calculation methods for velocities in the plane of the sky (e.g. optical flow) and \textit{Suzaku} data for the line-of-sight direction. Furthermore, we could characterize the CCO by detailed spectral analysis with better statistics than \cite{2020MNRAS.496.5052P}. We discuss a possible origin of the asymmetry of the ejecta structure and explosion of G350.1$-$0.3. We estimate the center of explosion in G350.1$-$0.3 for the first time and use this to argue for the NS kick scenario.

\subsection{Explosion center}

In this asymmetric object, the location of the center of the explosion is an extremely important aspect, especially in considering the relationship between the motion of CCO and the ejecta. Regarding the center of the explosion, previous studies such as \citep{2011ApJ...731...70L,2008ApJ...680L..37G} simply defined it as being in the middle of the bright region in X-rays or at the lowest point of the radio contour. But especially for CC SNRs, these explosion centers are poorly defined due to their asymmetric tendencies. Extrapolating the proper motion measured by image analysis can determine the kinematic center \citep{2006ApJ...645..283F,2017ApJ...845..167S}.

We estimated the time since the explosion by assuming that the ejecta had a common origin and drawing lines back to the presumed center adopting the velocities we measured and allowing the time since explosion to be a free parameter. Specifically, we calculated the time at which the variance of each origin point with respect to the presumed center was minimized. We investigated the case where the case (i) used two regions in the east and case (ii) used all five regions for which the maximum likelihood was used.
In the case (i), the center is at the position shown in figure 6, left panel, 460 years ago. The location of the explosion appears to be precisely determined, but the motion of the CCO cannot be explained. In this case, XMMU J172054.5$-$372652 would not be associated with the SNR. In the case (ii), the center is located at the position shown in figure 6 right panel, 655 years ago. The error circle is larger, but it explains the movement of the CCO. However, the degree of deceleration should vary from region to region, so there is that indeterminacy in the estimate of the explosion center \citep[e.g.,][]{sato2018x}. Our measurements assert that the center of the explosion lies closer to the CCO than previously thought.

\begin{figure*}[htb!]
 \centering
 \includegraphics[width=18cm]{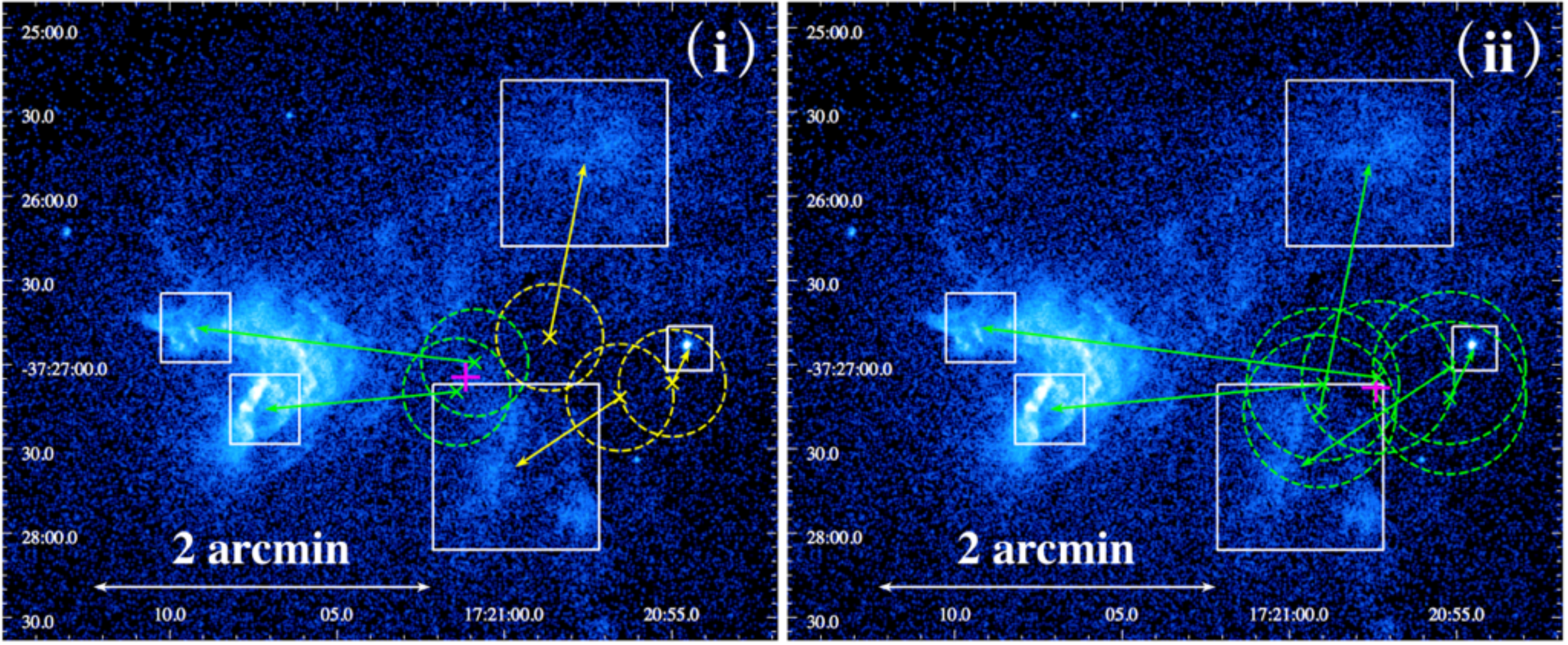}
 \caption{G350.1$-$0.3 image of 0.5--7.0 keV and explosive center using two regions in the east by the case (i) (left pannel) and explosive center using five regions by the case (ii) (right pannel). Green and yellow cross makers indicate the origins of each region when rewinding (i) 460 and (ii) 655 years. The green crosses are used to calculate the explosion center and the yellow crosses are not used. Pink cross makers indicate explosive centers determined by the case (i) and case the (ii), respectively. The dashed green circle shows the systematic error circle calculated by aligning the point sources.}
\end{figure*}

\vspace{0.3cm}

\subsection{Combined velocity field}

% combined analysis of radial and proper motion 

Combining the proper motions and line-of-sight velocities measured for the same region, we can construct a total velocity vector for each since the two speeds are orthogonal. The four regions except for the CCO in figure 1 and the six small boxes in figure 2b were used for this study. For the large regions shown in figure 1, velocity magnitudes of $\sim$5000 km s$^{-1}$ were obtained for East-1 and $\sim$3000 km s$^{-1}$ for East-2, North, and South. For the the entire six regions along the eastern ejecta shown in figure 2b, velocities of more than 4000 km s$^{-1}$ or close to 4000 km s$^{-1}$ were obtained. In particular, the ejecta motion was found to reach velocities of $\sim$5000 km s$^{-1}$ for box6, which is located at the lower end of the remnant.

We used the software to model the evolution of SNR \citep{Leahy_2017} to infer SNR properties from the 3D velocities obtained in this study.
We assumed a red supergiant as a progenitor and used a range of possible parameters of stellar wind mass loss and wind speed from \citep{2014ARA&A..52..487S}. The power-law indexes of the ISM and ejecta density profiles were fixed at 2 and 10 \citep{Matzner_1999}.
The velocity of contact discontinuity was assumed to be 0.75 times the velocity of blast-wave shock and the radius of contact discontinuity was assumed to be 0.75 times the radius of blast-wave shock \citep{Laming_2003}.
The measured velocity ($\sim$4800 km s$^{-1}$ for East-1) is assumed to be the contact discontinuity velocity, and the distance of 3.64 pc (by assuming 4.5 kpc) from the explosion center to the edge of the bright ejecta is considered to be the contact discontinuity radius. The age of the SNR was assumed to be 655 years. We searched for SNR properties consistent with these conditions.
Given the above assumptions, a model with ejecta energy of \(E_{\mathrm{ej}}\)=\(0.8-2 \times 10^{51} \mathrm{erg}\), ejecta mass of \(M_{\mathrm{ej}}\)=$4-15$\(M_{\odot}\) \citep{2014PASJ...66...68Y}, and stellar wind mass loss of \(10^{-6} M_{\odot} /\mathrm{yr} \) provides a reasonable interpretation for our measurements and the size observed by \textit{Chandra} in X-ray only in the area of East-1. However, when it comes to the South and North regions, it is impossible to explain the velocity with these properties. This implies that the spherically symmetric model cannot explain this object.

\subsection{Properties and Environment of XMMU J172054.5--372652}

The geometric center of the X-ray SNR and the XMMU J172054.5$-$372652 are about 1.5 arcmin apart, so it should be noted what is the relationship between the two parties. The typical values of the $N_{\mathrm{H}}$ for the ejecta and the $N_{\mathrm{H}}$ for the CCO are in agreement with values ranging from $3-4 \times 10^{22}\, \mathrm{cm}^{-2}$, consistent with the ejecta and the CCO having a common distance.

We explore the age of XMMU J172054.5$-$372652 from the theoretical cooling curves of NS \citep{2020MNRAS.496.5052P} and the surface temperature of the NS obtained by observations.
The fit gives the effective temperature \(\log T_{\mathrm{eff}} = 6.34_{-0.01}^{+0.02}\) K.
This result is reasonable given the cooling curve \citep{2020MNRAS.496.5052P}, with the age being around 655 years and a young NS.
In other words, XMMU J172054.5$-$372652 is most likely a neutron star associated with G350.1$-$0.3.

\subsection{NS kick}

G350.1$-$0.3 shows a highly asymmetric ejecta distribution, however, its kinematics has been unclear. Based on our ejecta velocity measurements in 3D space, we found most of the shocked ejecta at the eastern region in G350.1$-$0.3 are expanding with high velocities of $>$ 3500 km s$^{-1}$, which is comparable to that in the youngest galactic core-collapse SNR Cassiopeia A \citep{2010ApJ...725.2038D}, to the opposite direction from the NS position. As shown in figure 6, the estimated explosion centers in the two cases are both located to the east of the current NS position, so if the NS is receiving a kick, it should be receiving a kick to the west (opposite to the ejecta). The asymmetric ejecta/NS distribution and their kinematics in this remnant allow us to examine how asymmetric effects work for exploding massive stars.

 The ejecta expansion in the opposite direction to the NS kick supports a ``hydrodynamic kick'' mechanism.
 In theory, neutrino-driven convective mass flows \citep{1995ApJ...450..830B} and standing-accretion shock instability (SASI) sloshing motions \citep{2003ApJ...584..971B} could lead to large-scale asymmetries of the ejecta. The aspherical expanding ejecta gain radial momentum and its center of mass begins to shift away from the coordinate origin. Here the NS must receive the negative of the total momentum of the anisotropically expanding ejecta mass due to linear momentum conservation, which causes a NS kick in the opposite direction of the SN mass ejection. Recent hydrodynamic simulations of the neutrino-driven mechanism show such NS kicks up to more than 1000 km s$^{-1}$ \citep[e.g.,][]{2006A&A...457..963S,2010ApJ...725L.106W,2013A&A...552A.126W}. While the NS kick could also originate from an anisotropic emission of neutrinos \citep[``neutrino-driven kick'':][]{2006ApJS..163..335F}, however, the strongest mass ejection in the direction of the NS motion is predicted in this scenario, which is not suitable for G350.1$-$0.3.
 
The nucleosynthesis in G350.1$-$0.3 would be useful for understanding the asymmetric effects during the explosion. In the neutrino-driven mechanism, ``high-entropy bubbles'' in the process of forming large-scale asymmetries play an important role in the explosive nucleosynthesis \citep[e.g.,][]{2017ApJ...842...13W}. In the high-entropy nuclear burning region, the abundant $\alpha$ particles ($^{4}$He) are captured by heavy elements. Thus, the production of $\alpha$ elements (e.g., $^{44}$Ti) is enhanced in the high-entropy process. Thus, the strong asymmetry in the remnant implies a large production of $^{44}$Ti as in Cassiopeia A \citep[e.g.,][]{2014Natur.506..339G}, and the age of the remnant would encourage us to investigate the $^{44}$Ti emission\footnote{By assuming the age of 600 yr, the distance of 4.5 kpc, the total $^{44}$Ti mass of 10$^{-4}$ $M_\odot$, and the effective area at 68 keV of 70 cm$^2$, the expected flux is 3 $\times$ 10$^{-7}$ photons cm$^{-2}$ s$^{-1}$, which is almost comparable to the detection limit with the {\it NuSTAR} observation of $\sim$2 Ms \citep{2017ApJ...834...19G}.}. On the other hand, the $^{44}$Ti production depends not only on the entropy, but also on the electron fraction, $Y_{\rm e}$ (the average electron number per baryon) in the high-entropy region \citep[e.g.,][]{2010ApJS..191...66M,2013ApJ...774L...6W}. In particular, the amount of $^{44}$Ti in proton-rich high-entropy ejecta predicted in several theoretical simulations \citep[e.g.,][]{2018ApJ...852...40W} is much smaller than that in the neutron-rich ejecta \citep[][]{2017ApJ...842...13W}. Therefore, the neutron-rich environment is needed to create the abundant $^{44}$Ti in the remnant. The {\it \textit{Suzaku}} observation of G350.1$-$0.3 shows extremely high abundances of Ni compared to Fe \citep[ $Z_{\rm Ni}/Z_{\rm Fe} \approx$ 8:][]{2014PASJ...66...68Y}, which implies a suitable environment for the $^{44}$Ti production. As discussed here, the mass ratio among the iron-group elements and titanium could be a tracer of the nuclear burning conditions around the birth place of the NS. Further investigations for the element compositions in the remnant will be helpful to understand the supernova engine for driving CC SNe.

The asymmetric ejecta distribution and NS kick in G350.1$-$0.3 has not received much attention until now \citep[for example, the remnant is out of samples in][]{2017ApJ...844...84H,2018ApJ...856...18K}. We here succeeded in revealing the detailed kinematics of both ejecta and CCO in the remnant, which supports that the asymmetry in the kinematics of the remnant could originate from the neutrino-driven explosion and the strong hydrodynamic kick associated with the mechanism. Based on our results, it is more certain that the remnant can be one of the best samples in the future studies for discussing the origin of asymmetry in CC SNe such as Cassiopeia A.

\section{conclusion}
In this paper, we present the analysis of G350.1$-$0.3 and XMMU J172054.5--372652 of a bright source in X-rays with image and spectral analysis using data from two X-ray astronomy satellites.
There are three main points to consider.
\begin{itemize}
\item The velocities of the G350.1$-$0.3 ejecta are measured from three different image analyses in the plane of the sky due to the excellent spatial resolution of \textit{Chandra}, and in the line-of-sight direction consistent with the observed data of the two satellites. The three-dimensional velocities are as fast as $\sim$3000--5000 km s$^{-1}$, indicating that the G350.1$-$0.3 is a very young SNR.
\item For XMMU J172054.5$-$372652, the agreement with the column density of the SNR and the comparison between the surface temperature and the cooling curve suggest that the CCO originated from the same explosion as G350.1$-$0.3.
\item The explosion center, inferred from the current proper motion, is located east of the CCO, which supports the ``hydrodynamic kick'' mechanism depiction of the NS. However, measuring the kick of CCO by image analysis was difficult due to the large systematic error.
\end{itemize}
We took a closer look at the mysteries of the origin of this strange shaped SNR from the perspective of kinematics. More spanning monitoring observations are desired in the future to observe the NS kick and to study changes in the shape and velocity of the ejecta, which will lead to a better understanding of this explosion.

\section*{acknowledgements}
This work was supported by JSPS KAKENHI Grant Numbers 18H03722, 20H01941, and 20K20527.

\section*{Appendix. Velocities in the line-of-sight direction estimated from each of the observations in 2009 and 2018.}
Spectral analyses were performed separately for the 2009 observation and 2018 observations. The best-fit values of the redshift are obtained and summarized in table 4. The fitting model and parameter settings are the same as in Section 2.

\begin{deluxetable*}{ccccc}[ht!]
\tablecaption{Velocity measurements in the region shown in figure 1 for each of the observations in 2009 and 2018 and the merged ones.}
\tablehead{
  \colhead{region\tablenotemark{a}} & & \colhead{line-of-sight velocity(km s$^{-1}$)} & \\
  \cline{2-4}
  & \colhead{2009} & \colhead{2018} & \colhead{Two-epoch combined} 
}
\startdata
\hline
East-1  & $690 ^{+160}_{-160}$ & $1560 ^{+20}_{-20}$ & $1280 ^{+20}_{-20}$ \\
East-2   & $920 ^{+10}_{-10}$ & $1280 ^{+20}_{-10}$ & $1100 ^{+170}_{-190}$ \\
North  & $60 ^{+300}_{-410}$ & $660 ^{+130}_{-130}$ & $660 ^{+160}_{-110}$ \\
South  & $2740 ^{+10}_{-280}$ & $2540 ^{+40}_{-20}$ & $2570 ^{+140}_{-40}$ \\
\hline
\enddata
\tablenotetext{a}{East-1, East-2, North and South are indicated in figure 1.}
\label{tab:tab2}
\end{deluxetable*}


%\bibliography{bib.tex}{}

\begin{thebibliography}{}
\expandafter\ifx\csname natexlab\endcsname\relax\def\natexlab#1{#1}\fi
\providecommand{\url}[1]{\href{#1}{#1}}
\providecommand{\dodoi}[1]{doi:~\href{http://doi.org/#1}{\nolinkurl{#1}}}
\providecommand{\doeprint}[1]{\href{http://ascl.net/#1}{\nolinkurl{http://ascl.net/#1}}}
\providecommand{\doarXiv}[1]{\href{https://arxiv.org/abs/#1}{\nolinkurl{https://arxiv.org/abs/#1}}}

\bibitem[{{Anders} \& {Grevesse}(1989)}]{1989GeCoA..53..197A}
{Anders}, E., \& {Grevesse}, N. 1989, \gca, 53, 197,
  \dodoi{10.1016/0016-7037(89)90286-X}

\bibitem[{{Blondin} {et~al.}(2003){Blondin}, {Mezzacappa}, \&
  {DeMarino}}]{2003ApJ...584..971B}
{Blondin}, J.~M., {Mezzacappa}, A., \& {DeMarino}, C. 2003, \apj, 584, 971,
  \dodoi{10.1086/345812}

\bibitem[{{Borkowski} {et~al.}(2020){Borkowski}, {Miltich}, \&
  {Reynolds}}]{2020arXiv201112977B}
{Borkowski}, K.~J., {Miltich}, W., \& {Reynolds}, S.~P. 2020, arXiv e-prints,
  arXiv:2011.12977.
\newblock \doarXiv{2011.12977}

\bibitem[{{Burrows} {et~al.}(1995){Burrows}, {Hayes}, \&
  {Fryxell}}]{1995ApJ...450..830B}
{Burrows}, A., {Hayes}, J., \& {Fryxell}, B.~A. 1995, \apj, 450, 830,
  \dodoi{10.1086/176188}

\bibitem[{{DeLaney} {et~al.}(2010){DeLaney}, {Rudnick}, {Stage}, {Smith},
  {Isensee}, {Rho}, {Allen}, {Gomez}, {Kozasa}, {Reach}, {Davis}, \&
  {Houck}}]{2010ApJ...725.2038D}
{DeLaney}, T., {Rudnick}, L., {Stage}, M.~D., {et~al.} 2010, \apj, 725, 2038,
  \dodoi{10.1088/0004-637X/725/2/2038}

\bibitem[{Farneb{\"a}ck(2003)}]{farneback2003two}
Farneb{\"a}ck, G. 2003, in Scandinavian conference on Image analysis, Springer,
  363--370

\bibitem[{{Fesen} {et~al.}(2006){Fesen}, {Hammell}, {Morse}, {Chevalier},
  {Borkowski}, {Dopita}, {Gerardy}, {Lawrence}, {Raymond}, \& {van den
  Bergh}}]{2006ApJ...645..283F}
{Fesen}, R.~A., {Hammell}, M.~C., {Morse}, J., {et~al.} 2006, \apj, 645, 283,
  \dodoi{10.1086/504254}

\bibitem[{{Fryer} \& {Kusenko}(2006)}]{2006ApJS..163..335F}
{Fryer}, C.~L., \& {Kusenko}, A. 2006, \apjs, 163, 335, \dodoi{10.1086/500933}

\bibitem[{{Gaensler} {et~al.}(2008){Gaensler}, {Tanna}, {Slane}, {Brogan},
  {Gelfand}, {McClure-Griffiths}, {Camilo}, {Ng}, \&
  {Miller}}]{2008ApJ...680L..37G}
{Gaensler}, B.~M., {Tanna}, A., {Slane}, P.~O., {et~al.} 2008, \apjl, 680, L37,
  \dodoi{10.1086/589650}

\bibitem[{{Grefenstette} {et~al.}(2014){Grefenstette}, {Harrison}, {Boggs},
  {Reynolds}, {Fryer}, {Madsen}, {Wik}, {Zoglauer}, {Ellinger}, {Alexand er},
  {An}, {Barret}, {Christensen}, {Craig}, {Forster}, {Giommi}, {Hailey},
  {Hornstrup}, {Kaspi}, {Kitaguchi}, {Koglin}, {Mao}, {Miyasaka}, {Mori},
  {Perri}, {Pivovaroff}, {Puccetti}, {Rana}, {Stern}, {Westergaard}, \&
  {Zhang}}]{2014Natur.506..339G}
{Grefenstette}, B.~W., {Harrison}, F.~A., {Boggs}, S.~E., {et~al.} 2014, \nat,
  506, 339, \dodoi{10.1038/nature12997}

\bibitem[{{Grefenstette} {et~al.}(2017){Grefenstette}, {Fryer}, {Harrison},
  {Boggs}, {DeLaney}, {Laming}, {Reynolds}, {Alexander}, {Barret},
  {Christensen}, {Craig}, {Forster}, {Giommi}, {Hailey}, {Hornstrup},
  {Kitaguchi}, {Koglin}, {Lopez}, {Mao}, {Madsen}, {Miyasaka}, {Mori}, {Perri},
  {Pivovaroff}, {Puccetti}, {Rana}, {Stern}, {Westergaard}, {Wik}, {Zhang}, \&
  {Zoglauer}}]{2017ApJ...834...19G}
{Grefenstette}, B.~W., {Fryer}, C.~L., {Harrison}, F.~A., {et~al.} 2017, \apj,
  834, 19, \dodoi{10.3847/1538-4357/834/1/19}

\bibitem[{{Ho} \& {Heinke}(2009)}]{2009Natur.462...71H}
{Ho}, W. C.~G., \& {Heinke}, C.~O. 2009, \nat, 462, 71,
  \dodoi{10.1038/nature08525}

\bibitem[{{Holland-Ashford} {et~al.}(2017){Holland-Ashford}, {Lopez},
  {Auchettl}, {Temim}, \& {Ramirez-Ruiz}}]{2017ApJ...844...84H}
{Holland-Ashford}, T., {Lopez}, L.~A., {Auchettl}, K., {Temim}, T., \&
  {Ramirez-Ruiz}, E. 2017, \apj, 844, 84, \dodoi{10.3847/1538-4357/aa7a5c}

\bibitem[{{Hughes} {et~al.}(2000){Hughes}, {Rakowski}, {Burrows}, \&
  {Slane}}]{2000ApJ...528L.109H}
{Hughes}, J.~P., {Rakowski}, C.~E., {Burrows}, D.~N., \& {Slane}, P.~O. 2000,
  \apjl, 528, L109, \dodoi{10.1086/312438}

\bibitem[{{Katsuda} {et~al.}(2018){Katsuda}, {Morii}, {Janka},
  {Wongwathanarat}, {Nakamura}, {Kotake}, {Mori}, {M{\"u}ller}, {Takiwaki},
  {Tanaka}, {Tominaga}, \& {Tsunemi}}]{2018ApJ...856...18K}
{Katsuda}, S., {Morii}, M., {Janka}, H.-T., {et~al.} 2018, \apj, 856, 18,
  \dodoi{10.3847/1538-4357/aab092}

\bibitem[{{Koyama} {et~al.}(2007){Koyama}, {Tsunemi}, {Dotani}, {Bautz},
  {Hayashida}, {Tsuru}, {Matsumoto}, {Ogawara}, {Ricker}, {Doty}, {Kissel},
  {Foster}, {Nakajima}, {Yamaguchi}, {Mori}, {Sakano}, {Hamaguchi},
  {Nishiuchi}, {Miyata}, {Torii}, {Namiki}, {Katsuda}, {Matsuura}, {Miyauchi},
  {Anabuki}, {Tawa}, {Ozaki}, {Murakami}, {Maeda}, {Ichikawa}, {Prigozhin},
  {Boughan}, {Lamarr}, {Miller}, {Burke}, {Gregory}, {Pillsbury}, {Bamba},
  {Hiraga}, {Senda}, {Katayama}, {Kitamoto}, {Tsujimoto}, {Kohmura}, {Tsuboi},
  \& {Awaki}}]{2007PASJ...59S..23K}
{Koyama}, K., {Tsunemi}, H., {Dotani}, T., {et~al.} 2007, \pasj, 59, 23,
  \dodoi{10.1093/pasj/59.sp1.S23}

\bibitem[{Laming \& Hwang(2003)}]{Laming_2003}
Laming, J.~M., \& Hwang, U. 2003, The Astrophysical Journal, 597, 347,
  \dodoi{10.1086/378268}

\bibitem[{Leahy \& Williams(2017)}]{Leahy_2017}
Leahy, D.~A., \& Williams, J.~E. 2017, The Astronomical Journal, 153, 239,
  \dodoi{10.3847/1538-3881/aa6af6}

\bibitem[{{Lopez} {et~al.}(2009){Lopez}, {Ramirez-Ruiz}, {Badenes},
  {Huppenkothen}, {Jeltema}, \& {Pooley}}]{2009ApJ...706L.106L}
{Lopez}, L.~A., {Ramirez-Ruiz}, E., {Badenes}, C., {et~al.} 2009, \apjl, 706,
  L106, \dodoi{10.1088/0004-637X/706/1/L106}

\bibitem[{{Lovchinsky} {et~al.}(2011){Lovchinsky}, {Slane}, {Gaensler},
  {Hughes}, {Ng}, {Lazendic}, {Gelfand }, \& {Brogan}}]{2011ApJ...731...70L}
{Lovchinsky}, I., {Slane}, P., {Gaensler}, B.~M., {et~al.} 2011, \apj, 731, 70,
  \dodoi{10.1088/0004-637X/731/1/70}

\bibitem[{{Magkotsios} {et~al.}(2010){Magkotsios}, {Timmes}, {Hungerford},
  {Fryer}, {Young}, \& {Wiescher}}]{2010ApJS..191...66M}
{Magkotsios}, G., {Timmes}, F.~X., {Hungerford}, A.~L., {et~al.} 2010, \apjs,
  191, 66, \dodoi{10.1088/0067-0049/191/1/66}

\bibitem[{Matzner \& McKee(1999)}]{Matzner_1999}
Matzner, C.~D., \& McKee, C.~F. 1999, The Astrophysical Journal, 510, 379,
  \dodoi{10.1086/306571}

\bibitem[{{Millard} {et~al.}(2020){Millard}, {Bhalerao}, {Park}, {Sato},
  {Hughes}, {Slane}, {Patnaude}, {Burrows}, \& {Badenes}}]{2020ApJ...893...98M}
{Millard}, M.~J., {Bhalerao}, J., {Park}, S., {et~al.} 2020, \apj, 893, 98,
  \dodoi{10.3847/1538-4357/ab7db1}

\bibitem[{{Mitsuda} {et~al.}(2007){Mitsuda}, {Bautz}, {Inoue}, {Kelley},
  {Koyama}, {Kunieda}, {Makishima}, {Ogawara}, {Petre}, {Takahashi}, {Tsunemi},
  {White}, {Anabuki}, {Angelini}, {Arnaud}, {Awaki}, {Bamba}, {Boyce}, {Brown},
  {Chan}, {Cottam}, {Dotani}, {Doty}, {Ebisawa}, {Ezoe}, {Fabian}, {Figueroa},
  {Fujimoto}, {Fukazawa}, {Furusho}, {Furuzawa}, {Gendreau}, {Griffiths},
  {Haba}, {Hamaguchi}, {Harrus}, {Hasinger}, {Hatsukade}, {Hayashida}, {Henry},
  {Hiraga}, {Holt}, {Hornschemeier}, {Hughes}, {Hwang}, {Ishida}, {Ishisaki},
  {Isobe}, {Itoh}, {Iyomoto}, {Kahn}, {Kamae}, {Katagiri}, {Kataoka},
  {Katayama}, {Kawai}, {Kilbourne}, {Kinugasa}, {Kissel}, {Kitamoto}, {Kohama},
  {Kohmura}, {Kokubun}, {Kotani}, {Kotoku}, {Kubota}, {Madejski}, {Maeda},
  {Makino}, {Markowitz}, {Matsumoto}, {Matsumoto}, {Matsuoka}, {Matsushita},
  {McCammon}, {Mihara}, {Misaki}, {Miyata}, {Mizuno}, {Mori}, {Mori}, {Morii},
  {Moseley}, {Mukai}, {Murakami}, {Murakami}, {Mushotzky}, {Nagase}, {Namiki},
  {Negoro}, {Nakazawa}, {Nousek}, {Okajima}, {Ogasaka}, {Ohashi}, {Oshima},
  {Ota}, {Ozaki}, {Ozawa}, {Parmar}, {Pence}, {Porter}, {Reeves}, {Ricker},
  {Sakurai}, {Sanders}, {Senda}, {Serlemitsos}, {Shibata}, {Soong}, {Smith},
  {Suzuki}, {Szymkowiak}, {Takahashi}, {Tamagawa}, {Tamura}, {Tamura},
  {Tanaka}, {Tashiro}, {Tawara}, {Terada}, {Terashima}, {Tomida}, {Torii},
  {Tsuboi}, {Tsujimoto}, {Tsuru}, {Turner}, {Ueda}, {Ueno}, {Ueno}, {Uno},
  {Urata}, {Watanabe}, {Yamamoto}, {Yamaoka}, {Yamasaki}, {Yamashita},
  {Yamauchi}, {Yamauchi}, {Yaqoob}, {Yonetoku}, \&
  {Yoshida}}]{2007PASJ...59S...1M}
{Mitsuda}, K., {Bautz}, M., {Inoue}, H., {et~al.} 2007, \pasj, 59, S1,
  \dodoi{10.1093/pasj/59.sp1.S1}

\bibitem[{{Potekhin} {et~al.}(2020){Potekhin}, {Zyuzin}, {Yakovlev},
  {Beznogov}, \& {Shibanov}}]{2020MNRAS.496.5052P}
{Potekhin}, A.~Y., {Zyuzin}, D.~A., {Yakovlev}, D.~G., {Beznogov}, M.~V., \&
  {Shibanov}, Y.~A. 2020, \mnras, 496, 5052, \dodoi{10.1093/mnras/staa1871}

\bibitem[{{Sato} \& {Hughes}(2017)}]{2017ApJ...845..167S}
{Sato}, T., \& {Hughes}, J.~P. 2017, \apj, 845, 167,
  \dodoi{10.3847/1538-4357/aa8305}

\bibitem[{Sato {et~al.}(2018)Sato, Katsuda, Morii, Bamba, Hughes, Maeda,
  Ishida, \& Fraschetti}]{sato2018x}
Sato, T., Katsuda, S., Morii, M., {et~al.} 2018, The Astrophysical Journal,
  853, 46

\bibitem[{{Scheck} {et~al.}(2006){Scheck}, {Kifonidis}, {Janka}, \&
  {M{\"u}ller}}]{2006A&A...457..963S}
{Scheck}, L., {Kifonidis}, K., {Janka}, H.~T., \& {M{\"u}ller}, E. 2006, \aap,
  457, 963, \dodoi{10.1051/0004-6361:20064855}

\bibitem[{{Smith}(2014)}]{2014ARA&A..52..487S}
{Smith}, N. 2014, \araa, 52, 487, \dodoi{10.1146/annurev-astro-081913-040025}

\bibitem[{{Wanajo} {et~al.}(2013){Wanajo}, {Janka}, \&
  {M{\"u}ller}}]{2013ApJ...774L...6W}
{Wanajo}, S., {Janka}, H.-T., \& {M{\"u}ller}, B. 2013, \apjl, 774, L6,
  \dodoi{10.1088/2041-8205/774/1/L6}

\bibitem[{{Wanajo} {et~al.}(2018){Wanajo}, {M{\"u}ller}, {Janka}, \&
  {Heger}}]{2018ApJ...852...40W}
{Wanajo}, S., {M{\"u}ller}, B., {Janka}, H.-T., \& {Heger}, A. 2018, \apj, 852,
  40, \dodoi{10.3847/1538-4357/aa9d97}

\bibitem[{{Weisskopf} {et~al.}(2000){Weisskopf}, {Tananbaum}, {Van Speybroeck},
  \& {O'Dell}}]{2000SPIE.4012....2W}
{Weisskopf}, M.~C., {Tananbaum}, H.~D., {Van Speybroeck}, L.~P., \& {O'Dell},
  S.~L. 2000, in Society of Photo-Optical Instrumentation Engineers (SPIE)
  Conference Series, Vol. 4012, X-Ray Optics, Instruments, and Missions III,
  ed. J.~E. {Truemper} \& B.~{Aschenbach}, 2--16, \dodoi{10.1117/12.391545}

\bibitem[{{Wongwathanarat} {et~al.}(2010){Wongwathanarat}, {Janka}, \&
  {M{\"u}ller}}]{2010ApJ...725L.106W}
{Wongwathanarat}, A., {Janka}, H.-T., \& {M{\"u}ller}, E. 2010, \apjl, 725,
  L106, \dodoi{10.1088/2041-8205/725/1/L106}

\bibitem[{{Wongwathanarat} {et~al.}(2013){Wongwathanarat}, {Janka}, \&
  {M{\"u}ller}}]{2013A&A...552A.126W}
{Wongwathanarat}, A., {Janka}, H.~T., \& {M{\"u}ller}, E. 2013, \aap, 552,
  A126, \dodoi{10.1051/0004-6361/201220636}

\bibitem[{{Wongwathanarat} {et~al.}(2017){Wongwathanarat}, {Janka},
  {M{\"u}ller}, {Pllumbi}, \& {Wanajo}}]{2017ApJ...842...13W}
{Wongwathanarat}, A., {Janka}, H.-T., {M{\"u}ller}, E., {Pllumbi}, E., \&
  {Wanajo}, S. 2017, \apj, 842, 13, \dodoi{10.3847/1538-4357/aa72de}

\bibitem[{{Yasumi} {et~al.}(2014){Yasumi}, {Nobukawa}, {Nakashima}, {Uchida},
  {Sugawara}, {Tsuru}, {Tanaka}, \& {Koyama}}]{2014PASJ...66...68Y}
{Yasumi}, M., {Nobukawa}, M., {Nakashima}, S., {et~al.} 2014, \pasj, 66, 68,
  \dodoi{10.1093/pasj/psu043}

\end{thebibliography}
%\bibliographystyle{aasjournal}



\end{document}